\documentclass[usenatbib]{mn2e}
\usepackage{graphicx,natbib,times,amssymb}
\def\dnu{$\nu_{max}$}
\def\ddel{$\Delta\nu$}
\def\dhz{$\mu$Hz}
\def\dsm{$\mathrm{M}_{\odot}$}
\def\dmhef{$M_{\mathrm{HeF}}$}
\def\dmc{$M_{\mathrm{HeC}}$}

\title[Characteristics of solar-like oscillations of SRC stars]
{Characteristics of solar-like oscillations of secondary red clump stars}
\author[Yang et al.]{Wuming Yang$^{1,2}$\thanks{E-mail:
yangwuming@ynao.ac.cn; yang.wuming@yahoo.com.cn}, Xiangcun Meng$^{1}$,
Shaolan Bi$^{2}$, Zhijia Tian$^{2}$, Kang Liu$^{2}$,
\newauthor Tanda Li$^{2}$ and Zhongmu Li$^{3}$ \\
$^{1}$School of Physics and Chemistry, Henan Polytechnic University,
Jiaozuo 454000, Henan, China. \\
$^{2}$Department of Astronomy, Beijing Normal University, Beijing 100875,
China.\\
$^{3}$Institute for Astronomy and History of Science and Technology,
Dali University, Dali 671003, China.}

\begin{document}

\date{ }

\pagerange{\pageref{firstpage}--\pageref{lastpage}} \pubyear{2010}

\maketitle

\label{firstpage}

\begin{abstract}

We calculated the populations of core-helium-burning (CHeB) stars
and found that the secondary red clump (SRC) stars can form an SRC
peak in the distributions of the frequency of maximum seismic
amplitude (\dnu{}) and mean large-frequency separation (\ddel{}) of
CHeB stars when metallicity $Z \geq$ 0.02. The \dnu{} and \ddel{} of
CHeB stars are dependent not only on He core mass but on H-shell
burning. The SRC peak is composed of the CHeB stars with mass
roughly between the critical mass \dmhef{} and \dmhef{}$+0.2$ while
He core mass is between about 0.33 and 0.36 \dsm{}. The location of
the SRC peak can be affected by the mixing-length parameter
$\alpha$, metallicity $Z$, and overshooting parameter $\delta_{ov}$.
A decrease in $\alpha$ or increase in $Z$ or $\delta_{ov}$ leads to
a movement of the SRC peak towards a lower frequency. However, the
change in $Z$ and $\alpha$ only slightly affects the value of
\dmhef{} but the variation in $\delta_{ov}$ can significantly
affects the value of \dmhef{}. Thus the SRC peak might aid in
determining the value of \dmhef{} and calibrating $\delta_{ov}$. In
addition, the effects of convective acceleration of SRC stars and
the \dnu{} of `semi-degenerate' stars decreasing with mass result in
the appearance of a shoulder between about 40 and 50 \dhz{} in the
\dnu{} distribution. However, the convective acceleration of stars
with $M <$ \dmhef{} leads to the deficit in the \dnu{} distribution
between about 9 and 20 \dhz{}. Moreover, the value of the parameter
$b$ of the relation between \dnu{} and \ddel{} for the populations
with $M >$ \dmhef{} is obviously larger than that for the
populations with $M <$ \dmhef{}.

\end{abstract}


\section{Introduction}
Solar-like oscillations of giant stars were firstly confirmed in a
few stars \citep{fran02, barb04, ridd06}. And then the presence of
radial and non-radial solar-like oscillations in a large sample of
red giants was presented by \cite{ridd09}. Although only a very
limited number of modes ($l \leq$ 3) are likely to be observed in
solar-like oscillations due to geometrical cancellation effects, the
low-degree p-modes of the oscillations can penetrate deep into the
interior of stars \citep{dzie97}, and each low-degree p-mode carries
unique information about the stellar interior. Thus,
asteroseismology has the capability to probe the interior of stars
and to determine stellar fundamental parameters \citep{ulri86,
goug87, kjel95, chri02, yang07, yang10a, stel09a, stel09b, kall10a,
kall10b, moss10}.

However, for most of red giants, only the mean large-frequency
separation (\ddel{}) and the frequency of maximum seismic amplitude
(\dnu{}) have been obtained so far. The oscillation frequencies
\dnu{} and \ddel{} of a large sample of red giants were firstly
extracted by \cite{hekk09} from the data observed by the first
\emph{COnvection ROtation and planetary Transits (CoRoT)}
\citep{bagl06} 150-day long run in the direction of the Galactic
Centre (LRc01). They found that there is a dominant peak in the
distributions of \dnu{} and \ddel{} , respectively, and that there
is a tight relation between \dnu{} and \ddel{}, i.e.
\begin{equation}
    \Delta \nu \simeq a\nu_{max}^{b}, \label{eq1}
\end{equation}
where $a$ and $b$ are constant parameters. Using the data observed
by \emph{CoRoT} in the opposite direction of LRc01 (LRa01),
\cite{moss10} found similar results, except for the locations of the
dominant peaks and the values of the parameters $a$ and $b$. By
making use of the data observed by \emph{Kepler} mission
\citep{koch10}, \cite{hube10, hube11} and \cite{hekk11b} also
obtained very similar results. Moreover, the values of \dnu{} and
\ddel{} of red giants of clusters NGC 6811, NGC 6819, and NGC 6791
have also been extracted by \cite{stel10a} and \cite{hekk11a}. For
an ensemble of stars, the \dnu{} and \ddel{} can be applied to
investigate many interesting questions about the Galaxy or stellar
clusters \citep{migl09, migl12, migl11, yang10b, yang11a, hube10,
moss10, hekk11a, hekk11b, stel10a, stel10b, stel11}. These studies
advanced our understanding of the theory of stellar structure and
evolution.

\begin{figure}
     \includegraphics[width=8cm]{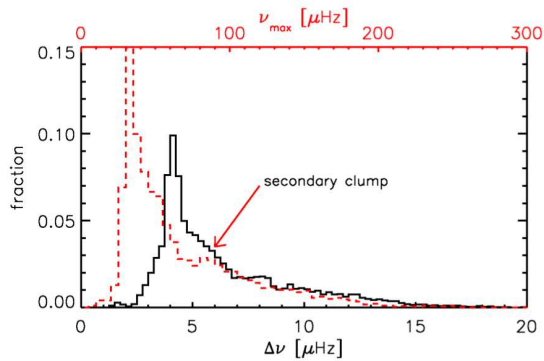}
     \centering
       \caption{The distribution of \dnu{} (red dashed line and top-axis) and
        \ddel{} (black solid line and bottom axis) obtained by \citet{hekk11b}.}
       \label{fig1}
   \end{figure}

Furthermore, it has been identified that the non-uniform
distributions of \dnu{} and \ddel{} of red giants result from
red-clump [core-helium-burning(CHeB)] stars \citep{migl09, moss10,
yang10b, hube10, hekk11b}. The dominant peak in the distributions of
\dnu{} and \ddel{}, which is located around 30 and 4 \dhz{} for
\dnu{} and \ddel{}, respectively, is mainly composed of the stars
that are close to the zero-age horizontal branch (ZAHB)
\citep{yang10b, yang11a}. Stars with \dnu{} $>$ 40 \dhz{} are mainly
red-clump stars with mass larger than 2.0 \dsm{} \citep{hube10}.
These stars could be attributed to the secondary red clump (SRC)
predicted by \cite{gira99} \citep{migl09, moss10, hube10, kall10b,
hekk11b}. Moreover, \cite{hekk11b} showed that the SRC stars could
become a hump in the \dnu{} distribution roughly between 70 and 100
\dhz{} [see Fig. \ref{fig1} or the Figure 2 in \cite{hekk11b}].
\cite{moss10} also showed that the distributions of \dnu{} and
\ddel{} observed by \emph{CoRoT} have a complex structure. The SRC
has been identified in the colour-magnitude diagrams of some
intermediate-age star clusters, such as NGC 419 in the Small
Magellanic Cloud \citep{gira09} and NGC 1751 in the Large Magellanic
Cloud \citep{rube11}. Although the origin of the SRC in clusters is
still an open question \citep{yang11b}, the SRC should be present in
all galactic fields containing stars with an age of about 1 Gyr
\citep{gira99}. The SRC is believed to be made of stars whose mass
is slightly more massive than the critical mass (\dmhef{}) that is
the maximum mass allowing helium flash to occur \citep{gira99,
yang11b}.

Stellar models with $M \gtrsim$ \dmhef{} can be generally evolved
from zero-age main sequence (ZAMS) to asymptotic giant branch (AGB)
without interruption. However, the models with $M <$ \dmhef{} are
hardly computed through the violent helium flash. Thus, the stars
with $M \gtrsim$ \dmhef{} should be more suitable to study
solar-like oscillations of CHeB stars and the effects of
convective-core overshooting and other physical processes on the
oscillations in details than the stars with $M <$ \dmhef{}. In
addition, the astroseismic detection and study on SRC stars may
provide additional constraints on the star-formation history in the
Galaxy and allow us to determine the value of \dmhef{}.

In this paper, we mainly focused on the global characteristics of solar-like
oscillations of SRC stars and on the effects of convective-core overshooting,
mixing-length parameter, and metallicity on the characteristics. The paper
is organized as follows. We show our stellar models and population-synthesis
method in section 2. We present the results in section 3 and discuss and
summarize them in section 4.

\section{Stellar models and population synthesis}
\subsection{Stellar models}
We used the stellar evolution code of \cite{egg71, egg72, egg73} to
calculate stellar evolution models. The code has been updated with
the recent input physics over the last three decades \citep{han94,
pols95, pols98}. The equation of state of \cite{eggl73} as modified
by \cite{pols95} and standard mixing-length theory are used in the
code. \textbf{The value of 2.0 for the mixing-length parameter 
($\alpha$) was calibrated against the Sun}. The convective-core
overshooting is treated as the prescription of \cite{schr97}. The
value of 0.12 for the overshooting parameter $\delta_{ov}$ is chosen
to match the properties of $\zeta$ Aurigae binaries which span a
mass range of 2.5 to 7.0 \dsm{} \citep{schr97}. The initial hydrogen
and helium abundances are assumed to be functions of the metallicity
as follows: $X = 0.76 - 3Z$ and $Y = 0.24 + 2Z$. The efficiency of
\cite{reim75} mass-loss rate is set to 0.4 for all models. The
opacity table for the metallicity is compiled by \cite{chen07} from
\cite{igle96} and \cite{alex94}. However, element diffusion for
helium and metals is not taken into account.

All models with $M \gtrsim$ \dmhef{} were evolved from ZAMS to AGB
stage without interruption. For the low-mass stars ($M <$ \dmhef{})
we were unable to compute through the He-flash. Instead, we
constructed ZAHB models using the prescription of \cite{pols98}.
These ZAHB models were evolved up to the AGB stage. It should be
noticed that some errors could be introduced in the post He-flash
evolution of these constructed ZAHB models \citep{pols98}.

In addition, for a star given mass, radius and effective temperature,
the theoretical \dnu{} and \ddel{} were calculated by using scaling
equations \citep{brow91, kjel95}
\begin{equation}
    \nu_{max}=3050 \frac{M/M_{\odot}}{(R/R_{\odot})^{2}
    \sqrt{T_{eff}/5777 K}}\mathrm{\mu Hz}\,,
    \label{eq2}
\end{equation}
   and
\begin{equation}
    \Delta \nu=134.6 \frac{(M/M_{\odot})^{1/2}}{(R/R_{\odot})^{3/2}}
     \mathrm{\mu Hz}\,. \label{eq3}
\end{equation}
Here, the value of 134.6 \dhz{} for $\Delta\nu_{\odot}$ is obtained
from the GONG data \citep{yang09}. The accuracy of these estimates
is good to within 5\% \citep{stel09a}.

\subsection{Stellar population synthesis}
\label{sps}
In order to obtain the distributions of \dnu{} and \ddel{} of SRC stars,
we calculated single-star stellar population (SSP). Stellar samples are
generated by the Monte Carlo simulation. The basic assumptions for the
simulations are as follows. (i) Star formation rate (SFR) is assumed to
be a constant. (ii) The age-metallicity relation is taken from \cite{roch00}
or a constant metallicity is assumed for simplicity. (iii) The lognormal
initial mass function (IMF) of \cite{chab01} is adopted.

\section{Calculation results}
\subsection{The peak of SRC stars}
In order to obtain the \dnu{} and \ddel{} of the populations of SRC
stars, we firstly generated a ZAMS sample with mass between 1.80 and
5.0 \dsm{} according to the IMF of \cite{chab01}. Then we computed
the evolutions of the models in the sample from high to low mass
until calculations were interrupted by a violent He-flash at the tip
of first giant branch (FGB). For a given $Z$, we define \dmhef{} as
the mass for which the \dnu{} at the beginning of quiescent helium
burning [`horizontal branch (HB)'. Hereafter, the `HB' also refers to
the similar evolutionary stage of models with $M >$ \dmhef{} in the
Hertzsprung-Russell diagram.] reaches a maximum and He-core
mass (\dmc{}) reaches a minimum. The luminosity of the model with $M =$
\dmhef{} at the FGB tip almost reaches a minimum. The value of the
He-core mass is about 0.33 \dsm{} for $Z =$ 0.02 and about 0.34 \dsm{}
for $Z =$ 0.004. The values of \dmhef{} in Table \ref{tab1} for
different metallicities are estimated from the tracks calculated at
an interval less than 0.005 \dsm{}. These values are almost not
affected by the mixing-length parameter $\alpha$ but can be
significantly affected by overshooting parameter $\delta_{ov}$. The
stars with mass slightly less than \dmhef{} (for $Z =$ 0.02, the mass
is between about 1.70 and 2.01 \dsm{}) ignite helium under
semi-degenerate conditions at core masses between $\sim$ 0.33 and
0.45 \dsm{} (\emph{hereafter referred to as semi-degenerate stars}).
However, for the stars with lower mass, helium is ignited in the
degenerate core with mass about 0.45 \dsm{} (\emph{hereafter referred
to as degenerate stars}) \citep{pols98}.

\begin{table}
\centering \caption{The values of the critical mass \dmhef{} (in
\dsm{}) estimated from our calculations with $\alpha$ = 2.0 and
$\delta_{ov}$ = 0.12.}
\begin{tabular}{lllllll}
  \hline
   Z & 0.05 & 0.04 & 0.03 & 0.02 & 0.01 & 0.004 \\
 \hline
 \dmhef{} & 1.96 & 1.99 & 2.01 & 2.01 & 1.93 & 1.83 \\
\hline
\end{tabular}
\label{tab1}
\end{table}
   \begin{figure}
     \includegraphics[width=6cm, angle=-90]{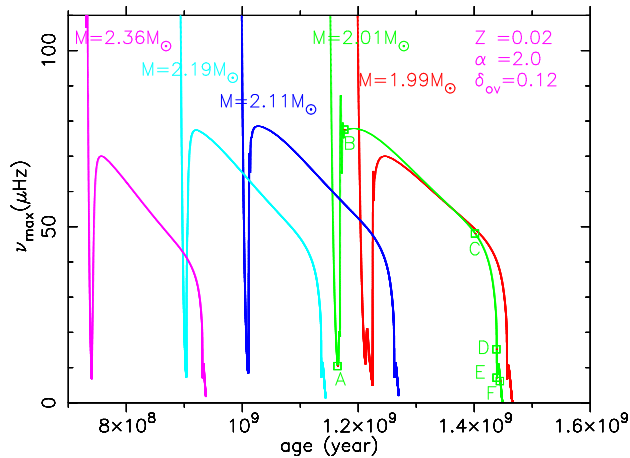}
     \centering
     \caption{The evolutions of \dnu{} of CHeB stars with different masses.
     These models were evolved from ZAMS to AGB. For Z = 0.02 and
     $\delta_{ov} = 0.12$, the critical mass \dmhef{} is 2.01 \dsm{}.
     But the minimum mass that can be evolved from ZAMS to AGB without
     interruption by He-flash is 1.99 \dsm{} in our calculations.
     Point A corresponds to the FGB tip. Point B indicates the state that
     the star just reaches the `HB'. Point C shows the state that a
     convective core has formed and central He abundance decreases
     to about 0.18. Point D represents the state that the central He is
     completely depleted. Point E shows the state that central temperature
     and the rate of H-shell burning reaches a maximum, respectively.
     Point F shows the state that the H-shell is reignited, ie. the
     terminal of the early AGB.}
       \label{fig2}
   \end{figure}
   \begin{figure}
     \includegraphics[width=6cm, angle=-90]{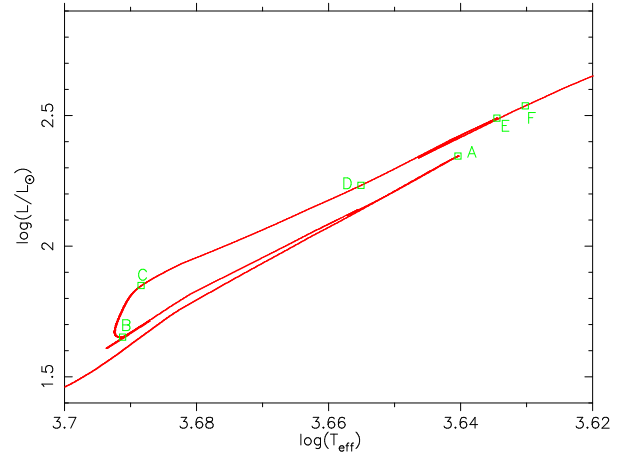}
     \centering
       \caption{The Hertzsprung-Russell diagram of the model with
       $M =$ 2.01 \dsm{} and $Z =$ 0.02. The points A, B, C, D, E,
       and F correspond to the points in Fig. \ref{fig2}.}
       \label{fig3}
   \end{figure}

   \begin{figure}
     \includegraphics[width=9cm, angle=-90]{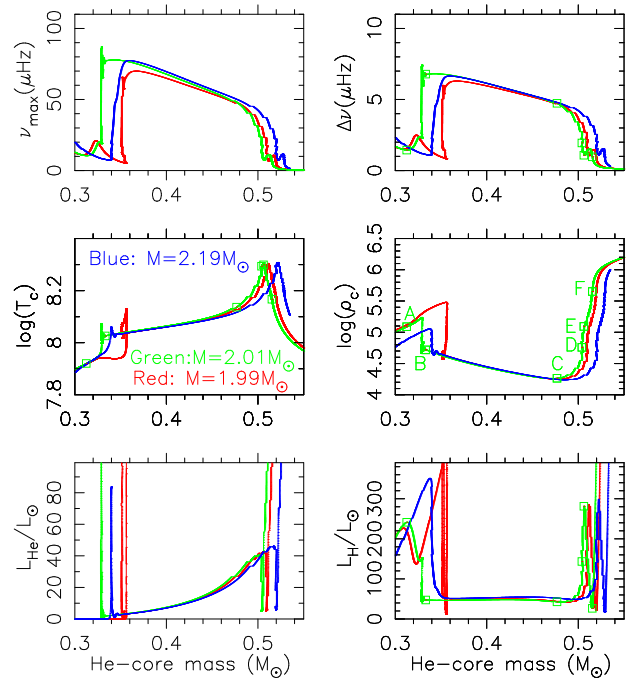}
     \centering
       \caption{The \dnu{}, \ddel{}, central temperature $T_{c}$,
       central density $\rho_{c}$, He-burning luminosity
       $L_{\mathrm{He}}$, and H-burning luminosity $L_{\mathrm{H}}$
       as a function of He-core mass.}
       \label{fig4}
   \end{figure}

The evolutions of \dnu{} of CHeB stars with the same $Z$, $\alpha$,
and $\delta_{ov}$ but different masses are shown in Fig. \ref{fig2}.
The \dnu{}, \ddel{}, central temperature $T_{c}$, central density
$\rho_{c}$, He-burning luminosity $L_{\mathrm{He}}$, and H-burning
luminosity $L_{\mathrm{H}}$ of three models are shown in Fig.
\ref{fig4} as a function of He-core mass. Several specifically
evolutionary states are marked in the Figs. \ref{fig2}, \ref{fig3}
and \ref{fig4}. The point A on the track of $M = 2.01$ \dsm{}
corresponds to the FGB tip in the Hertzsprung-Russell diagram (see
Fig. \ref{fig3}), while the point B represents the beginning of the
`HB'. From point B to C, helium burning proceeds quietly in the He
core, the central He abundance $Y_{c}$ decreases gradually to about 0.18,
and a convective core has formed in the stars. Then owing to mixing
in the core, there is a sudden depletion of He fuel over a large
region, which leads to a rapid contraction of the core, increase in
central temperature and expansion of the radius of the star on a
thermal time-scale. This results in that the star evolves quickly from
$Y_{c} \approx$ 0.18 to $Y_{c} \approx$ 0 (point D) and that the
\dnu{} decreases rapidly. We labeled this evolutionary phase as
`convective acceleration'. From point D, the star evolves through
the early AGB. As the He abundance in central regions goes to zero,
the He-exhausted core contracts and heats up while the H-rich
envelope expands and cools. Cooling in the H-rich envelope is so
effective that the H-shell burning begins to extinguish \citep{chio92}
and the envelope recontracts (point E), which leads to the increase
in \dnu{} and \ddel{}. Eventually the contraction is prevented
by the efficient He-shell burning. Then the stars begin to expand and
their \dnu{} and \ddel{} begin to decrease. The H shell is reignited at
point F. From then on H-shell burning dominates energy production. Thus
there is a hook between point E and F of the evolutionary track of
\dnu{}, and the stars take a long time from E to F. As a consequence,
there might be a peak (AGB peak) in the \dnu{} distribution around
10 \dhz{}.

When stars arrive at the `HB', their \dnu{} and \ddel{} are mainly
dependent on their He-core mass and H-shell burning. When the He-core
mass is larger than about 0.36 \dsm{}, the \dnu{} and \ddel{} obviously
decrease with the increase in He-core mass. For the semi-degenerate
stars, when they reach the `HB', their He-core masses increase fast
with decreasing initial mass. Thus the values of \dnu{} and \ddel{}
of these stars decrease rapidly with decreasing mass. For example,
when the mass decreases from 2.01 to 1.99 \dsm{}, the He-core mass
increases from about 0.33 to 0.37 \dsm{} and the value of \dnu{}
decreases from about 78 to 70 \dhz{} (see Fig. \ref{fig2}).
Moreover, for a given mass star, when the energy production of
H-shell burning arrives at a minimum, its \dnu{} almost reaches
a maximum. When the He-core masse of the model with $M =$ 2.01
\dsm{} increases from about 0.33 to 0.36 \dsm{}, the energy production
of CHeB increases from about 2.0 to 3.6 L$_\odot$, but the energy
from H-shell burning decreases from about 47.5 to 46.4 L$_\odot$
and reaches a minimum. The total energy production is almost constant.
The changes in the luminosity, radius, and mean density
of this model are insignificant in this stage. As a consequence, its
\dnu{} and \ddel{} are almost unchanged. For the star with $M =$
2.19 \dsm{}, when it reaches the `HB', its He-core mass is about
0.36 \dsm{}, and its energy production of H-shell burning reaches a
minimum. Its \dnu{} reaches a maximum at this time and is almost
equal to that of model with $M =$ 2.01 \dsm{} at \dmc{}
$\simeq$ 0.36 \dsm{} (see Fig. \ref{fig4}). When the stars with
mass between 2.01 and 2.19 \dsm{} arrive at the `HB', their
He-core masses are between about 0.33 and 0.36 \dsm{}. Their energy
productions of H-shell burning also reach a minimum as their
\dmc{} $\simeq$ 0.36 \dsm{}. The \dnu{} of these models is similar
to that of model with $M =$ 2.01 \dsm{}. For the stars with $M >$
2.19 \dsm{}, when they arrive at the `HB', their He-core masses are
larger than 0.36 \dsm{}. The more massive the star, the larger the
He-core mass, thus the smaller the \dnu{} and \ddel{}. For example,
when the mass increases from 2.19 to 2.36 \dsm{}, the value of \dnu{}
decreases from 77.5 to 70.0 \dhz{} (see Fig. \ref{fig2}).

   \begin{figure}
     \includegraphics[width=6cm, angle=-90]{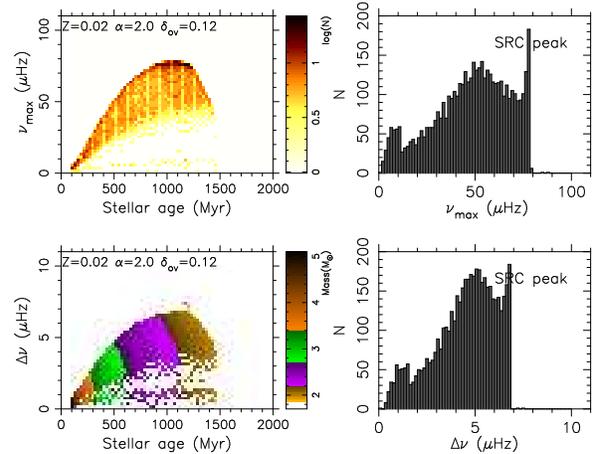}
     \centering
       \caption{The distributions of \dnu{} and \ddel{} of the CHeB
        stars with $M \gtrsim$ \dmhef{} as a function
        of stellar age and the histograms of the \dnu{} and \ddel{} of
        these stars.}
       \label{fig5}
   \end{figure}
Figure \ref{fig5} shows the distributions of \dnu{} and \ddel{} of
the CHeB populations with mass between 5.0 and 1.99 \dsm{} and $Z =$
0.02 as a function of stellar age and the histograms of the \dnu{} and
\ddel{}. The convective acceleration leads to the absence of the
SRC stars with \dnu{} $<$ 50 \dhz{} in the \dnu{} distribution. The
CHeB stars with mass between about 2.19 and 2.01 \dsm{} form an SRC
peak at about 75-78 \dhz{} for \dnu{} and at about 6.5-6.8 \dhz{}
for \ddel{}. If the stars with mass less than 1.99 \dsm{} were
included, the peak at about 50 \dhz{} for \dnu{} and about 5
\dhz{} for \ddel{} might disappear, which depends on the properties
of semi-degenerate stars. However, the SRC peak can not be affected
by the stars because the \dnu{} and \ddel{} of semi-degenerate stars
decrease rapidly with decreasing mass (see the comparision in Fig.
\ref{fig10}). Moreover, the peak located at about 10 \dhz{} for
\dnu{} and 1 \dhz{} for \ddel{} is mainly composed of the CHeB
stars with $M \gtrsim$ 4.0 \dsm{} and the stars with lower mass
but undergone convective acceleration. The stars
with mass between about 2.19 and 2.01 \dsm{} have an almost equal
\dnu{} when they arrive at the `HB'. However, for the stars with $M >$
2.19 ($M <$ 2.01) \dsm{}, the value of \dnu{} increases (decreases)
with decreasing mass. This might provide us an opportunity to
distinguish the SRC stars from others in the asteroseismical
observations of a large sample of stars.

\subsection{The effect of metallicity on the distributions of \dnu{}
and \ddel{} of SRC stars}

   \begin{figure}
     \includegraphics[width=7cm, angle=-90]{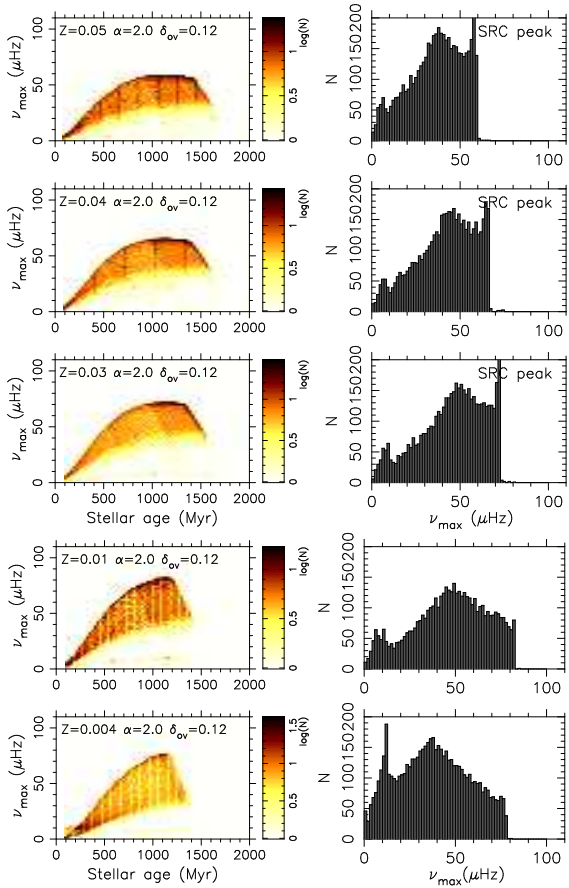}
     \includegraphics[width=4.7cm, angle=-90]{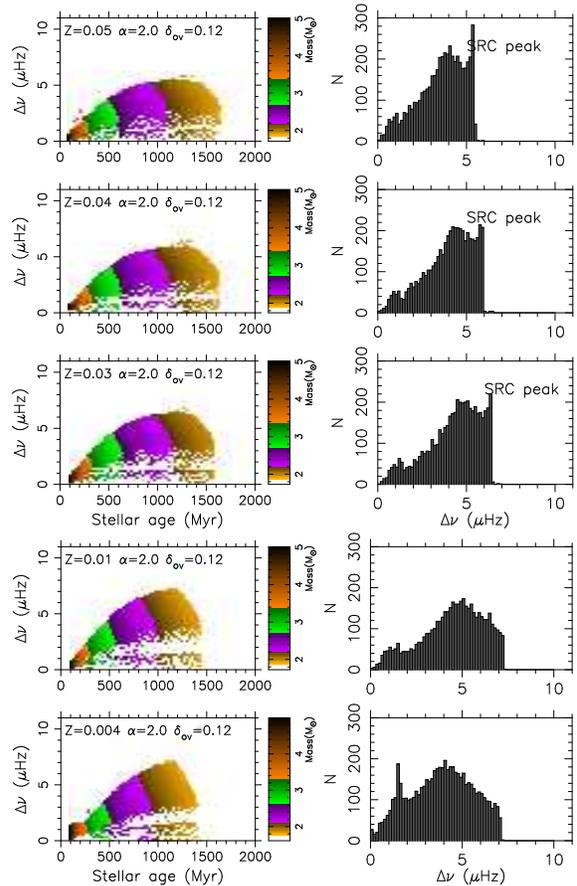}
     \centering
      \caption{Same as Fig. \ref{fig5} but for different metallicities.}
      \label{fig6}
   \end{figure}
   \begin{figure}
     \includegraphics[width=7cm, angle=-90]{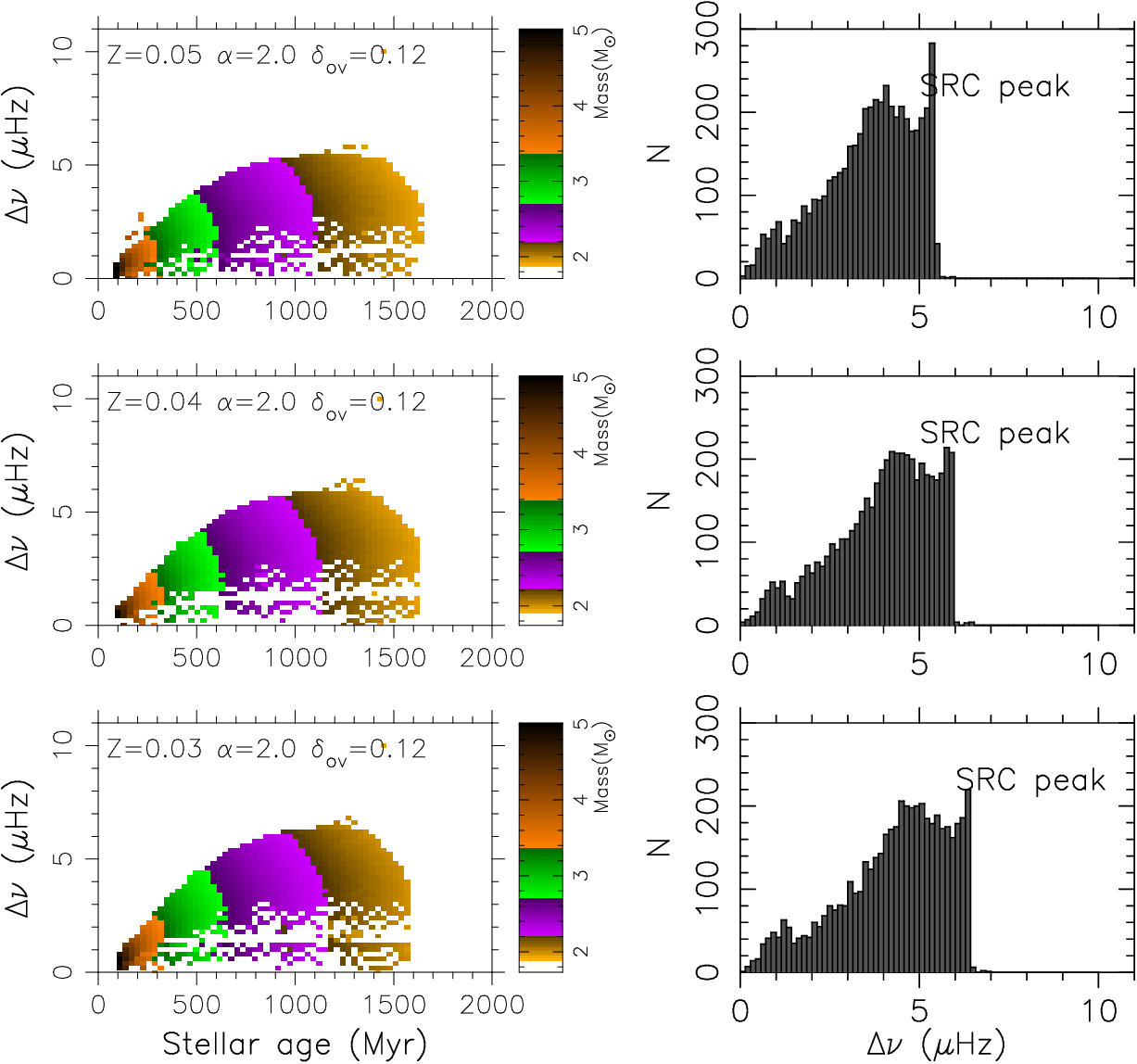}
     \includegraphics[width=4.7cm, angle=-90]{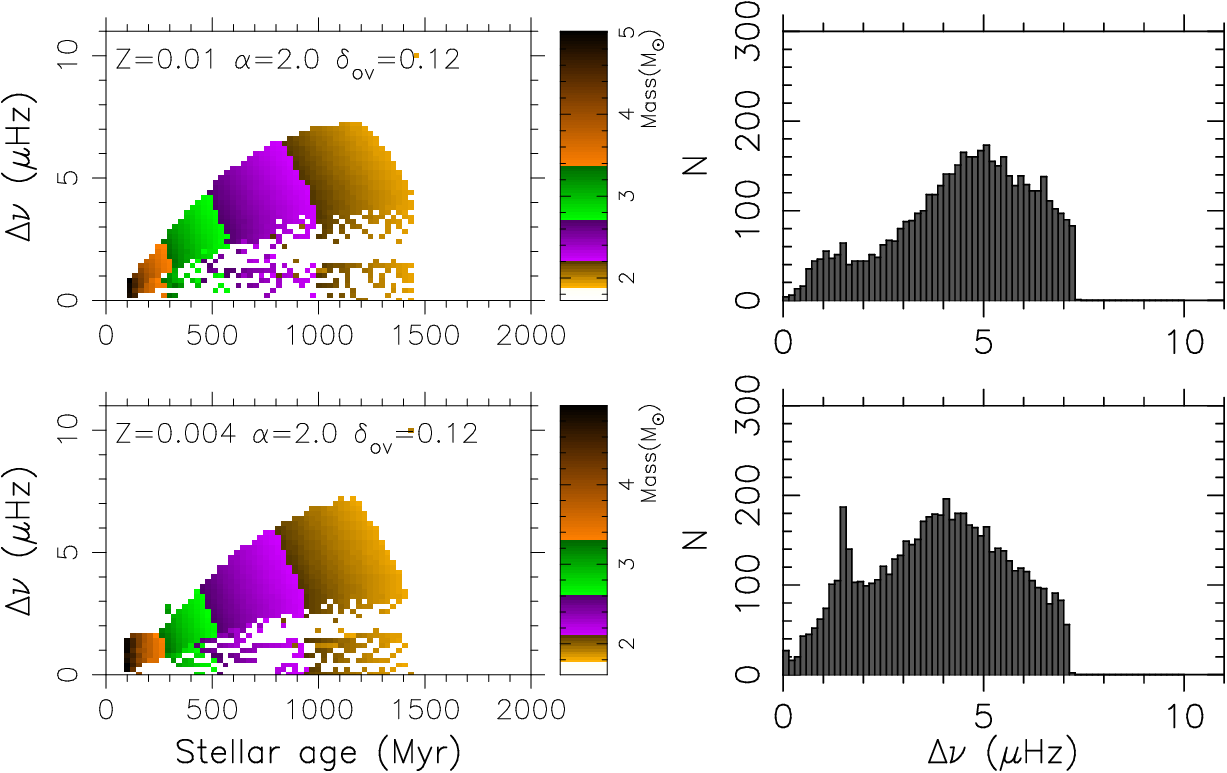}
     \centering
      \caption{Same as Fig. \ref{fig5} but for different metallicities.}
      \label{fig7}
   \end{figure}
We calculated the CHeB populations with the same $\alpha$ and
$\delta_{ov}$ but different metallicities. The distributions of
\dnu{} and \ddel{} of the populations are shown in Figs. \ref{fig6}
and \ref{fig7}. Our calculations show that the SRC peak is present
in the histograms of \dnu{} and \ddel{} of populations with $Z
\gtrsim$ 0.02. However, when the metallicity increases from 0.02 to
0.05, the value of \dmhef{} decreases from 2.01 to 1.96 \dsm{} and
the location of the SRC peak moves from about 78 to 58 \dhz{} for
\dnu{}. When the stars with $M =$ \dmhef{} and $Z \gtrsim$ 0.02
arrive at the `HB', they have almost the same He core, but
their radii and the energy production of H-shell burning increase
with increasing metallicity, i.e. the higher the metallicity the
greater the energy production of H-shell burning and the less the
contraction of H-rich envelope when the stars evolve from the tip
of the FGB to the `HB'. Thus, the mean density of critical mass
models decreases with increasing metallicity. Hence the values
of the \dnu{} and \ddel{} of the models decrease with increasing
metallicity. Therefore, the frequency of the location of the SRC
peak decreases with increasing metallicity.

\begin{figure}
     \includegraphics[width=6cm, angle=-90]{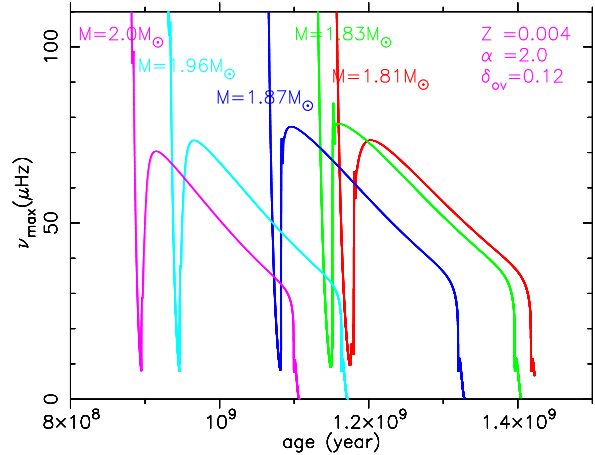}
     \centering
     \caption{Same as Fig. \ref{fig2} but for $Z =$ 0.004. For $Z =$
      0.004 and $\delta_{ov} = 0.12$, the critical mass \dmhef{} is
      1.83 \dsm{}.}
     \label{fig8}
   \end{figure}

\begin{figure}
     \includegraphics[width=9cm, angle=-90]{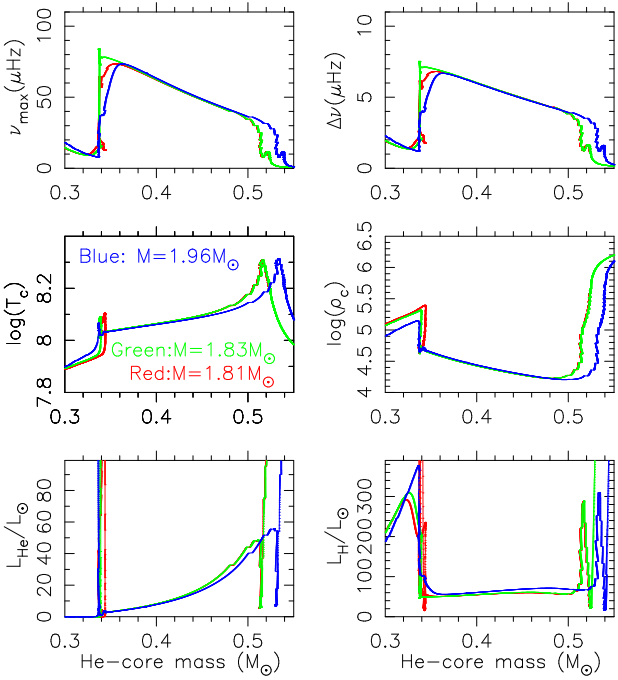}
     \centering
       \caption{\textbf{Same as Fig. \ref{fig4}} but for $Z =$ 0.004.
       For $Z =$ 0.004 and $\delta_{ov} = 0.12$, the critical
       mass \dmhef{} is 1.83 \dsm{}.}
       \label{fig9}
   \end{figure}

However, the SRC peak does not appear in the histograms of \dnu{}
and \ddel{} of the populations with $Z =$ 0.01 and 0.004. Figure
\ref{fig8} shows the evolutions of \dnu{} of CHeB stars with the
same metallicity (0.004) but different masses. The maximum of \dnu{}
of the CHeB stars with $M >$ \dmhef{} (1.83 \dsm{}) decreases with
increasing stellar mass, which is different from the result of
populations with $Z =$ 0.02. The He-core mass of the model with
$M =$ 1.83 \dsm{} and $Z =$ 0.004 is about 0.34 \dsm{} when the
star arrive at the `HB', the energy production of H-shell burning
reaches a minimum at the same time (see Fig. \ref{fig9}).
The burning rates of central helium and H-shell increase
with the increase in the He-core mass, which makes the star
to expand. Thus its \dnu{} and \ddel{} decrease with the increase
in the He-core mass. When a star with $M >$ 1.83 \dsm{} arrives
at the `HB', its He-core mass is larger than 0.34 \dsm{},
its energy production of H-shell burning reaches a minimum at
the same time, and its \dnu{} arrives at a maximum.
The more massive the stellar mass the bigger the He-core mass, the
smaller the \dnu{}. Thus the SRC peak can not be formed in the
histogram of \dnu{} of populations with $Z =$ 0.004. The
characteristics of \dnu{} and \ddel{} of populations with
$Z =$ 0.01 are similar to those of populations with $Z =$ 0.004.


\subsection{The effects of the mixing-length parameter, degenerate and
semi-degenerate stars on the SRC peak}

When $\alpha =$ 2.0 and $\delta_{ov} =$ 0.12, the ages of the stars
with mass slightly larger than \dmhef{} are located roughly between
1.0 and 1.3 Gyr as the stars reach the `HB'. According to the
age-metallicity relation of \cite{roch00}, the metallicity of the
stars with an age of about 1.2 Gyr is around 0.04. For the populations
with $Z = 0.04$, $\alpha = 2.0$, and $\delta_{ov} = 0.12$, the SRC peak
is located at about 63-67 \dhz{} in the \dnu{} distribution and at around
5.7-6.0 \dhz{} in the \ddel{} distribution. We note that, in the figures
4 and 5 of \cite{moss10}, there seems to be a bump at around 65 \dhz{}
in the \dnu{} distribution and at about 5.5-6.0 \dhz{} in the \ddel{}
distribution, which may be corresponding to our SRC peak. However,
the secondary clump found by \cite{hekk11b} in the \dnu{} distribution
is located roughly between 70 and 100 \dhz{}. Increasing $\alpha$
can decrease stellar radius, i.e. can increase \dnu{} and \ddel{}. Thus
we calculated the populations with a bigger $\alpha$.

\begin{figure}
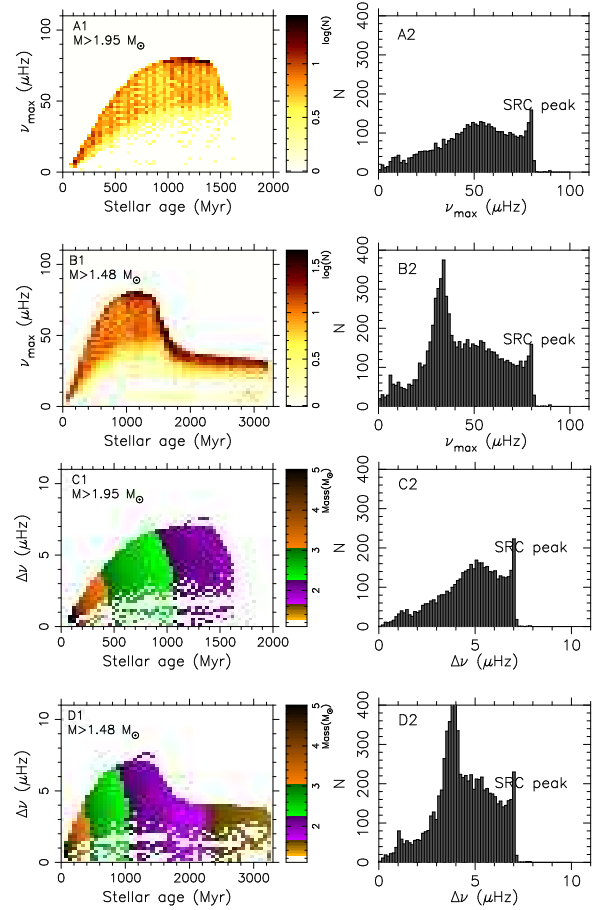

     \includegraphics[width=6cm, angle=-90]{pgz4n.ps}
     \includegraphics[width=6cm, angle=-90]{pgz4d.ps}
     \centering
       \caption{Same as Fig. \ref{fig5} but for the populations with $Z = 0.04$,
        $\delta_{ov} = 0.12$, and $\alpha = 2.7$. The value of \dmhef{} for
        $Z = 0.04$, $\delta_{ov} = 0.12$, and $\alpha = 2.7$ is 1.98 \dsm{}.
        The populations with $M >$ 1.48 \dsm{} include the degenerate and
        semi-degenerate stars.}
       \label{fig10}
   \end{figure}

Figure \ref{fig10} shows the distributions of \dnu{} and \ddel{} of
the CHeB populations with $Z = 0.04$, $\delta_{ov} = 0.12$, and
$\alpha = 2.7$. The value of \dmhef{} is about 1.98 \dsm{} for this
set of parameters. The change in $\alpha$ hardly affects the value
of \dmhef{}. However the location of SRC peak moves from about 65 to
80 \dhz{} for \dnu{} and from 5.8 to 7.0 \dhz{} for \ddel{} when the
value of $\alpha$ increases from 2.0 to 2.7. Red giant stars have a
deep convective envelope. Increasing $\alpha$ leads to an increase
in the efficiency of convective energy transport, which can cause
the more contraction of the convective envelope. The larger the
$\alpha$ the more the contraction. However, the change in $\alpha$
barely affects the nuclear reaction in stellar interior. Therefore
the variation of the mixing-length parameter has almost no influence
on the age and luminosity but can change the mean density of models.
Hence, the values of \dnu{} and \ddel{} increase with increasing
$\alpha$.

In order to investigate the effect of stars with $M <$ \dmhef{}
on the SRC peak, the panels B's and D's of Fig. \ref{fig10} show
the distributions of \dnu{} and \ddel{} of populations with $M >$
1.48 \dsm{}. These panels show that the SRC peak can not be affected
by the degenerate and semi-degenerate stars. The dominant peak of
\dnu{} and \ddel{} of these populations is located at about 33 and 4
\dhz{}, respectively, and is mainly caused by the stars with age
larger than about 2.0 Gyr. The value of \dnu{} and \ddel{} of the
CHeB stars with $M <$ 1.48 \dsm{} is almost less than 40 and 4
\dhz{}, respectively. These stars can not affect the distributions
of \dnu{} larger than 40 \dhz{} and \ddel{} larger than 4.0 \dhz{}
unless their mass-loss rate is very high in red giant stage
\citep{yang10b}, but can slightly affect the location of the
dominant peak.

Moreover, panel B2 of Fig. \ref{fig10} shows that there is a
shoulder in the \dnu{} histogram roughly between 40 and 50 \dhz{}.
The convective acceleration of stars with $M \gtrsim$ \dmhef{}
leads to the absence of stars with \dnu{} less than about 50 \dhz{}
and age between about 0.9 and 1.5 Gyr in the panels A1 and B1 of
Fig. \ref{fig10}. The panel A2 of Fig. \ref{fig10} shows that the
number of stars with $M \gtrsim$ \dmhef{} increase when \dnu{}
decreases from 70 to about 50 \dhz{}, and then decrease with
the decrease in \dnu{}. We marked this peak in the \dnu{} histogram
around 50 \dhz{} as convective acceleration peak. The value of
about 50 \dhz{} is determined by the convective acceleration
of stars with $M \gtrsim$ \dmhef{}. However, the value of \dnu{}
of semi-degenerate stars decreases with decreasing stellar mass.
In addition, according to the IMF, the lower the stellar mass
the greater the number of stars. Thus the number of stars with
$M <$ \dmhef{} increase with the decrease in \dnu{}.
The shoulder derives from that the decrease in the number of
stars with $M >$ \dmhef{} is just counteracted by the increase
in the number of stars with $M <$ \dmhef{} between
about 40 and 50 \dhz{}.

\subsection{The effects of convective core overshooting on the SRC peaks}

\begin{figure}
     \includegraphics[width=7cm, angle=-90]{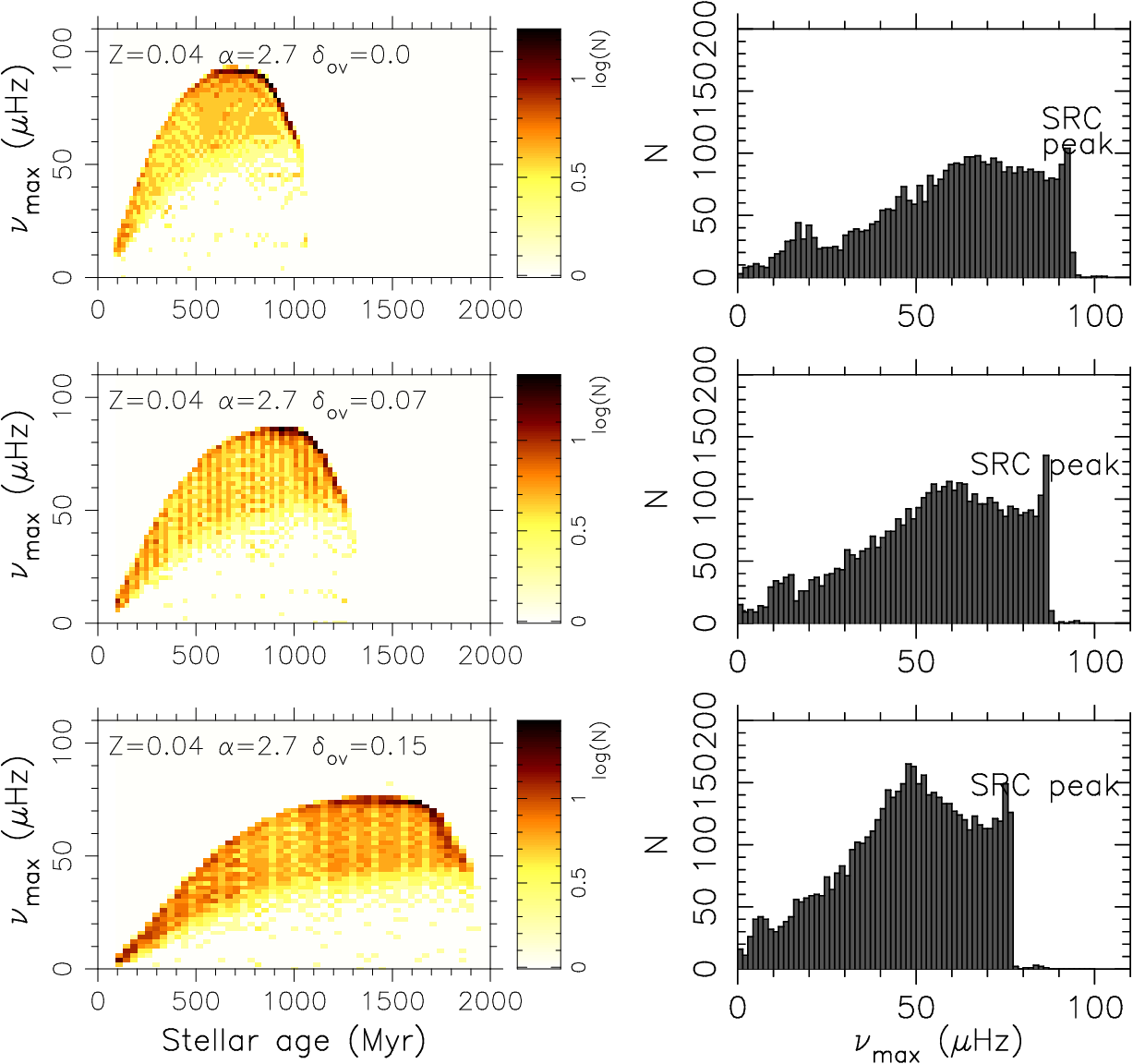}
     \centering
       \caption{Same as Fig. \ref{fig10} but for the populations
        with different
        $\delta_{ov}$.}
       \label{fig11}
   \end{figure}

\begin{figure}
     \includegraphics[width=7cm, angle=-90]{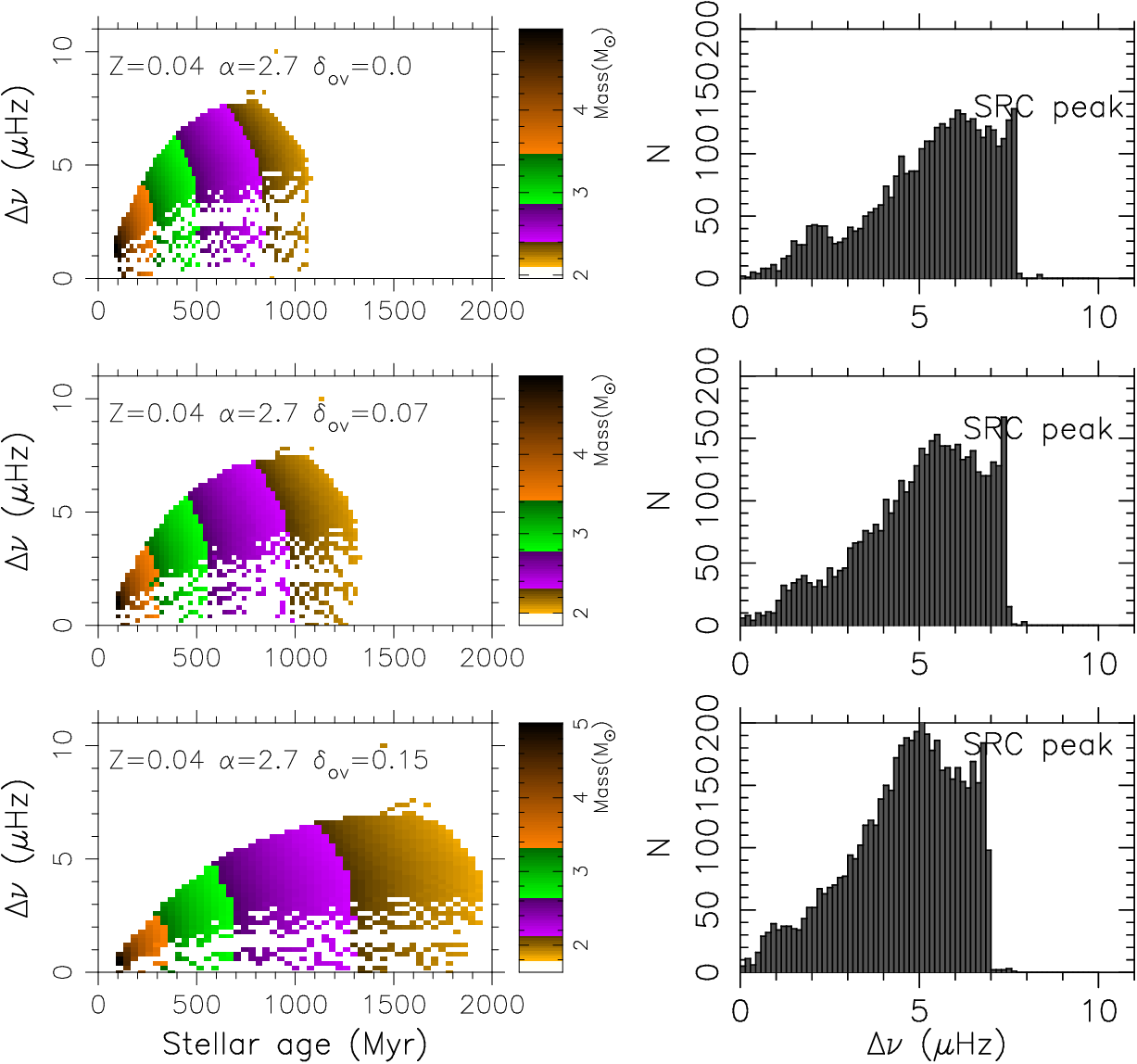}
     \centering
       \caption{Same as Fig. \ref{fig10} but for the populations
        with different
        $\delta_{ov}$.}
       \label{fig12}
   \end{figure}

\begin{figure}
     \includegraphics[width=6cm, angle=-90]{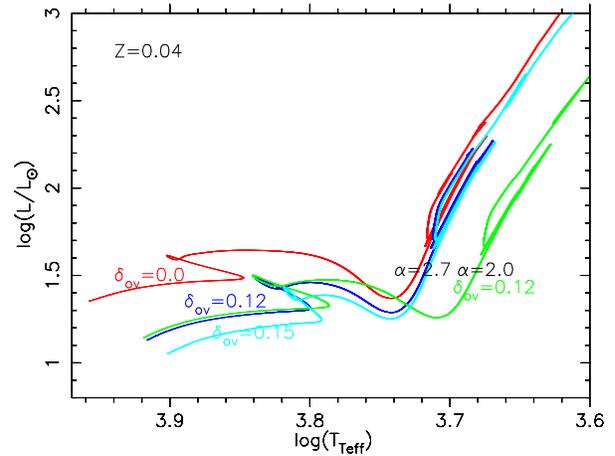}
     \centering
       \caption{The evolutionary tracks of the models with $M$ = \dmhef{}.
       These models have the same metallicity ($Z$ = 0.04) but different
       $\alpha$ and $\delta_{ov}$ which are labeled on their tracks. }
       \label{fig13}
   \end{figure}

Figs.\ref{fig11} and \ref{fig12} show the distributions of \dnu{}
and \ddel{} of populations with the same $Z$ and $\alpha$ but
different $\delta_{ov}$. For the populations with $Z$ = 0.04 and
$\alpha$ = 2.7, when the value of the $\delta_{ov}$ decreases from
0.15 to 0.0, the value of \dmhef{} increases from 1.89 to 2.24 \dsm{}
and the location of the SRC peak moves from about 76 to 91 \dhz{}
for \dnu{} and from about 6.7 to 7.6 \dhz{} for \ddel{}. Moreover,
the age of the stars composed the SRC peak is between about 0.6 and
0.8 Gyr for $\delta_{ov} =$ 0.0 but the age is between around 1.2
and 1.6 Gyr for $\delta_{ov} =$ 0.15. This is mainly caused by the
change in \dmhef{}. We listed the values of the \dmhef{} for different
$\delta_{ov}$ in Table \ref{tab2} and plotted the evolutionary tracks
of the models with the critical mass in Fig. \ref{fig13}.
An increase in $\delta_{ov}$ prolongs the lifetime of
core H burning by feeding more H-rich material into the core, which
can enhance the He core mass left behind and strongly change the
global characteristics of the following giant stages \citep{schr97}.
Thus the value of \dmhef{} decreases with increasing $\delta_{ov}$.
When the models with these critical masses reach the `HB', they have
almost the same He-core mass, central temperature, central density,
central pressure, and nuclear energy production. However, the larger
the critical mass the smaller the stellar radius, and the more
the contraction of H-rich envelope when the stars evolve from the FGB
tip into the `HB', i.e. the higher the mean density. Thus the decrease
in $\delta_{ov}$ leads to a movement of the location of SRC peak
towards a higher frequency. However, the change in $\delta_{ov}$
has almost no influence on the effective temperature of SRC stars (see
Fig. \ref{fig13}).

\begin{table}
\centering \caption{The values of the critical mass \dmhef{} (in
\dsm{}) estimated from our calculations with $\alpha$ = 2.7 and $Z$
= 0.04. }
\begin{tabular}{lllllll}
  \hline
   $\delta_{ov}$ & 0.15 & 0.12 & 0.10 & 0.07 & 0.00 \\
 \hline
 \dmhef{}        & 1.89 & 1.98 & 2.03 & 2.11 & 2.24 \\
\hline
\end{tabular}
\label{tab2}
\end{table}

\subsection{Parameters $a$ and $b$ }
   \begin{figure}
     \includegraphics[width=6.5cm, angle=-90]{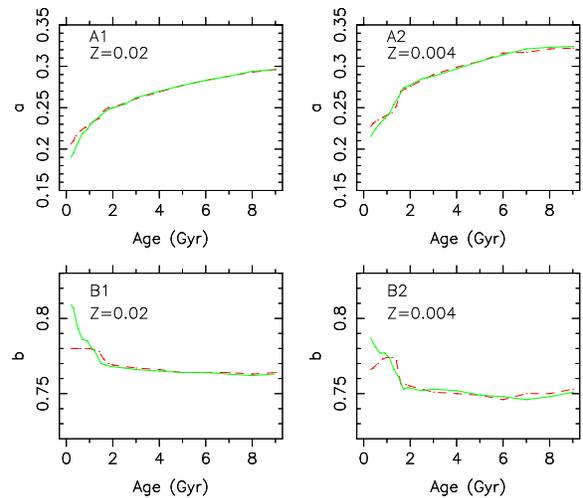}
     \centering
       \caption{The changes of parameters $a$ and $b$ with the age of
        populations. The dashed lines show the results of the SSP
        without spread in age, while the solid lines indicate those of
        the SSP with a spread in age of 200 Myr.}
       \label{fig14}
   \end{figure}

Figure \ref{fig14} shows the parameters $a$ and $b$ as a function of
population age. For the populations with $Z$ = 0.02, the value of
$a$ increases from about 0.2 to 0.3 when the age of populations
increases from about 0.2 to 9.0 Gyr. However, the value of parameter
$b$ is about 0.78 and almost unchanged when age $<$ 1.2 Gyr and is
around 0.764 but slightly decreases with age when age $>$ 2.0 Gyr.
When 1.2 Gyr $<$ age $<$ 2.0 Gyr, the value of $b$ decreases with
age. Recent observations show that the intermediate-age star
clusters may be composed of stars with a non-uniform age [a spread
in age of about 200-300 Myr \citep{mack07, milo09}]. Our
calculations show that a spread in age of 200 Myr can significantly
affect the parameters $a$ and $b$ of the populations with age less
than about 2 Gyr but hardly affects those of the populations with
age $>$ 2 Gyr. The spread leads to an increase in $b$ but a decrease
in $a$ for the populations with age less than about 1.2 Gyr and a
decrease in $b$ for the populations with age between about 1.2 and
2.0 Gyr. For the populations with the spread in age, the value of
$b$ decreases fast from about 0.81 to 0.77 when the age of the
populations increases from 0.2 to 2.0 Gyr. Moreover, our
calculations show that the characteristics of $a$ and $b$ of
populations with $Z$ = 0.03, 0.04 and 0.05 are almost the same as
those of populations with $Z$ = 0.02. However, the value of $a$ of
populations with $Z$ = 0.004 is larger than that of populations with
$Z$ = 0.02. But the value of $b$ of populations with $Z$ = 0.004 is
smaller than that of populations with $Z$ = 0.02. When the age of
populations is larger than about 2 Gyr, the value of $b$ for the
populations with 0.02 $\leq Z \leq$ 0.04 is located between about
0.77 and 0.76; however, that for populations with $Z$ = 0.004 is
located between about 0.76 and 0.75. The value of $b$ of populations
with $Z =$ 0.01 is located between that of populations with $Z =$
0.02 and populations with $Z =$ 0.004. Furthermore, our calculations
show that a change in the mixing-length parameter does not
significantly affect the parameters $a$ and $b$. However, the
decrease in $\delta_{ov}$ can lead to a small increase in $b$ and
decrease in $a$ for the populations with $M \gtrsim$ \dmhef{}.

The value of parameters $a$ and $b$ is about 0.20 and 0.79 for NGC
6811 ($\sim$1.0 Gyr), 0.24 and 0.77 for NGC 6819 ($\sim$2.5 Gyr),
and 0.28 and 0.75 for NGC 6791 ($\sim$8.5 Gyr), respectively
\citep{hekk11a}. The characteristics of the $a$ and $b$ of
populations with the spread in age of 200 Myr are more consistent
with these obtained by \cite{hekk11a} than those of populations with
a uniform age.

\subsection{A synthesis result }
   \begin{figure}
     \includegraphics[width=6cm, angle=-90]{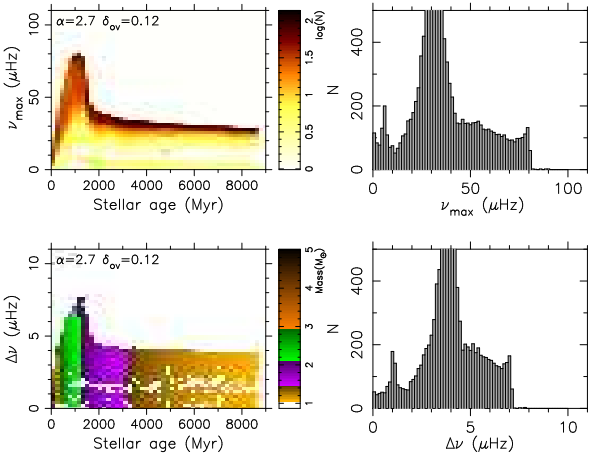}
     \centering
       \caption{Same as Fig. \ref{fig4} but for the populations with age
       between 0 and 9 Gyr.}
       \label{fig15}
   \end{figure}
We computed the stellar populations with $\alpha$ = 2.7,
$\delta_{ov}$ = 0.12, and $Z$ = 0.004, 0.01, 0.02, 0.03, 0.04 and
0.05. However, the metallicity of the populations in the Galaxy is
believed to change with age. We estimated the metallicity from the
age-metallicity relation of \cite{roch00}. By interpolating, we
obtained the CHeB populations with age between 0 and 9 Gyr and
metallicity changing with age. The distributions of \dnu{} and
\ddel{} of the populations are shown in Fig. \ref{fig15}. The
dominant peak is located about 30 and 3.9 \dhz{} in the \dnu{} and
\ddel{} distribution, respectively. The SRC peak is situated at
around 78 and 7 \dhz{} in the \dnu{} and \ddel{} distribution,
respectively. The shoulder is located between about 43 and 54 \dhz{}
in the \dnu{} distribution. However, there is not a significant
shoulder in the \ddel{} distribution. Moreover, there is a deficit
in the \dnu{} distribution between about 9 and 20 \dhz{} and
in the \ddel{} distribution between around 1.2 and 2.5 \dhz{},
which is caused by the convective acceleration of CHeB stars.
The peak in the \dnu{} histogram about 7 \dhz{} and in the \ddel{}
histogram around 1 \dhz{} is mainly composed of the CHeB stars
with $M \gtrsim$ 4.0 \dsm{} and the early AGB stars.
This peak exists in the distributions of \dnu{} and \ddel{}
observed by the \emph{CoRoT} \citep{moss10, hube10}.
The deficit in the \dnu{} histogram between about 10 and 20 \dhz{}
and in the \ddel{} histogram between around 1.5 and 2.5 \dhz{}
observed by \emph{CoRoT} \citep{moss10} might be partly
caused by the convective acceleration.

\section{Discussions and Conclusions}

The value of \dmhef{} obtained from our definition is slightly
different from that obtained from the definition of \cite{pols98}
who defines \dmhef{} as the mass for which the luminosity at the FGB
tip reaches a minimum and the core mass at He ignition is about 0.33
\dsm{}. According to \cite{pols98} definition, we obtained the value
of \dmhef{} is 1.994 \dsm{} for $Z =$ 0.02, $\alpha =$ 2.0, and
$\delta_{ov} =$ 0.12. The value is consistent with that obtained by
\cite{pols98}. Our calculations show that the maximum of \dnu{} of
the model with $M =$ 1.994 \dsm{} is about 76 \dhz{} on the `HB',
which is less than the value of 78 \dhz{} of the model with $M =$
2.01 \dsm{}.

The SRC peak can appear in the distributions of \dnu{} and \ddel{}
of populations with $Z \geq$ 0.02 in our simulations. The location
of the SRC peak can be affected by the parameters $Z, \alpha$, and
$\delta_{ov}$. An increase in $\alpha$ or decrease in $Z$ or
$\delta_{ov}$ can lead to an increase in the frequency of the
location of the SRC peak. The change in $Z$ or $\alpha$ can
significantly affect the effective temperature of the populations of
SRC stars rather than the age of the populations and the value of
\dmhef{} for $Z \geq$ 0.02. However, the variation in $\delta_{ov}$
can significantly affect the value of \dmhef{} and the age of the
populations rather than the effective temperature.
According to the formulas (\ref{eq2}) and (\ref{eq3}), the mass of a
star can be estimated from its \dnu{}, \ddel{} and effective
temperature $T_{eff}$. If the frequency of the location of the SRC
peak can be obtained from observations, which may aid in determining
the value of \dmhef{} and calibrating the overshooting parameter
$\delta_{ov}$. Moreover, The SRC peak is mainly composed of CHeB
stars with mass between about \dmhef{} and \dmhef{} + 0.2 whose
He-core masses are between about 0.33 and 0.36 \dsm{}.
Because the mass range of the SRC peak stars is very limited,
the SRC peak in the histograms of the \dnu{} and \ddel{} may aid
in determining the star-formation rate in the Galaxy or star
clusters.

The SRC peak is considerable in our simulations. Comparing with the
dominant peak, it is, however, insignificant. If the observed SRC
stars have different $Z, \alpha$ and $\delta_{ov}$, the SRC peak
might become inconsiderable.

The appearance of the shoulder in the histogram of \dnu{} is related
to the convective acceleration of stars with mass $\gtrsim$ \dmhef{}
and the \dnu{} of semi-degenerate stars decreasing with mass.
The degenerate and semi-degenerate zero-age models are always constructed
from a non-degenerate model by mass loss. The \dnu{} and \ddel{} of
these models are mainly dependent on the He-core mass of the zero-age
models rather than other parameters. We do not know the exact value of
the He-core mass of stars with $M <$ \dmhef{} when they reach the `HB'.
A small change in the He-core mass can lead to an obvious variation in
the \dnu{} of semi-degenerate stars. Thus the distributions of \dnu{} and
\ddel{} of semi-degenerate stars may be affected by the construction method
of the ZAHB models. If the \dnu{} of semi-degenerate stars
decreases with mass too fast, the increase in the number of
semi-degenerate stars with decreasing \dnu{} can not counteract the
decrease in the number of stars with $M >$ \dmhef{} between about
40 and 50 \dhz{}, the convective acceleration peak will appear in the
simulated histogram of \dnu{} of populations including semi-degenerate
and degenerate stars, while the upper boundary of the shoulder location
will move to a lower frequency, i.e. the shoulder might be replaced
by a narrower shoulder plus a convective acceleration peak. The
frequency of the upper boundary of the shoulder or the location of
convective acceleration peak is mainly determined by the frequency
where the convective acceleration of SRC stars begins. The value of
this frequency can be affected by the parameters $Z, \alpha$
and $\delta_{ov}$. For example, for populations with $Z =$ 0.04, the
value is around 54 \dhz{} for $\alpha =$ 2.7 but is about 47
\dhz{} for $\alpha =$ 2.0. However, the frequency of the lower
boundary of the shoulder is mainly determined by the degenerate stars.
The effect of $Z, \alpha$ and $\delta_{ov}$ on the lower boundary is
harder to study by evolving stellar models than that on the upper boundary.
A detailed statistical study on the red ginat stars with observed \dnu{}
$>$ 40 \dhz{} would provide additional constraints on the theory of
structure and evolution of semi-degenerate and SRC stars.

The SRC and semi-degenerate stars with the same \dnu{} have
different \ddel{}, i.e. the values of parameters $a$ and $b$ for
SRC stars are different from those for semi-degenerate stars,
which leads to a small difference between the \dnu{} and \ddel{}
distributions. For example, the shoulder is less significant in
the \ddel{} distribution than that in the \dnu{} distribution.

The parameter $a$ of CHeB populations increases with the age of
populations in the range of about 0.20 $-$ 0.33. The value of the
parameter $b$ for populations with $M >$ \dmhef{} (age $\lesssim$
1.1 Gyr) is obviously larger than that for the populations
with $M <$ \dmhef{}. The value of $b$ for old populations is
approximately equal. In general, a spread in the age of populations
leads to a decrease in $a$ and increase in $b$ for the populations
with $M \gtrsim$ \dmhef{} but results in a small increase in $a$
and decrease in $b$ for the populations with $M <$ \dmhef{}.

When the star with $M =$ \dmhef{} and $Z =$ 0.02 evolved into the
`HB', its He-core mass is about 0.33 \dsm{}. When the He-core mass
increases from about 0.33 to 0.36 \dsm{}, the energy production of
He burning increases slightly but that of H-shell burning decreases
and reaches a minimum, the total energy production is almost constant.
The \dnu{} and \ddel{} of this model is almost unchanged in this phase.
When the stars with mass slightly larger than \dmhef{} (roughly between
\dmhef{} and \dmhef{} + 0.2) and $Z =$ 0.02 arrive at the `HB',
their He-core masses are between 0.33 and 0.36
\dsm{} and their \dnu{} and \ddel{} are almost equal. Our
simulations show that these stars can form an SRC peak in the
histograms of \dnu{} and \ddel{} of the CHeB populations. The SRC
peak can also appear in the histograms of \dnu{} and \ddel{} of the
CHeB populations with $Z =$ 0.03, 0.04, and 0.05. However, it does
not exist in the histograms of \dnu{} and \ddel{} of the CHeB
populations with $Z =$ 0.01 and 0.004. The SRC peak is present in the
histograms of \dnu{} and \ddel{} of the CHeB populations calculated
according to the age-metallicity relation of \cite{roch00} too.
The SRC peak location can be affected by the mixing-lenght
parameter $\alpha$, core overshooting parameter $\delta_{ov}$ and
metallicity $Z$. An increase in $\alpha$ or decrease in
$\delta_{ov}$ or $Z$ leads to a movement of the peak location
towards a higher frequency. The change in $\delta_{ov}$ can
significantly affect the value of \dmhef{} but hardly affects the
effective temperature of the CHeB populations, which is different
from the effects of $\alpha$ and $Z$. The SRC peak would aid in
determining the value of \dmhef{} and calibrating the overshooting
parameter $\delta_{ov}$. Moreover, the convective acceleration in
the stars with $M \gtrsim$ \dmhef{} and the \dnu{} of semi-degenerate
stars decreasing with mass result in the appearance
of a shoulder in the \dnu{} distribution. The convective
acceleration of stars with $M <$ \dmhef{} leads to the deficit
in the \dnu{} histogram between about 9 and 20 \dhz{} and
in the \ddel{} histogram between about 1.5 and 2.5 \dhz{}.

\section*{Acknowledgments}
This work was supported by China Postdoctoral Science Foundation
funded project 20100480222, the Ministry of Science and Technology
of the People's republic of China through grant 2007CB815406,
the NSFC though grants 10773003, 10933002, 11003003, 10963001, and
the Project of Science and Technology from the Ministry of
Education (211102)
.

\end{document}